# Widening Perspectives: The Intellectual and Social Benefits of Astrobiology (Regardless of Whether Extraterrestrial Life is Discovered or Not)[1]

**I.A. Crawford**

Department of Earth and Planetary Sciences, Birkbeck College, University of London, Malet Street, London, WC1E 7HX
Email: i.crawford@bbk.ac.uk

**Abstract**

Astrobiology is usually defined as the study of the origin, evolution, distribution, and future of life in the universe. As such it is inherently interdisciplinary and cannot help but engender a worldview infused by cosmic and evolutionary perspectives. Both these attributes of the study of astrobiology are, and will increasingly prove to be, beneficial to society regardless of whether extraterrestrial life is discovered or not.

## Introduction

The principal objective of astrobiology as a scientific discipline is to understand the origin, evolution, distribution, and future of life in the universe. Clearly, the 'holy grail' of astrobiology would be the actual discovery of life elsewhere in the universe, and such a discovery would have profound scientific, and very likely also philosophical and societal, implications (although the nature of these implications will doubtless depend on the nature and location of the extraterrestrial life discovered). Needless-to-say, there will also be significant scientific and philosophical implications if extraterrestrial life is *not* discovered, despite ever more sophisticated searches for it. Given that the outcome of the search for life in the universe is still uncertain, in this essay I wish to address the intellectual and social benefits of the *study* of astrobiology, regardless of whether or not the search is ultimately successful.

---

[1] To comply with the publisher's terms and conditions, this is the submitted version of a manuscript now accepted for publication in the *International Journal of Astrobiology*. The published peer-reviewed version, possibly with minor changes, will be available at https://www.cambridge.org/core/journals/international-journal-of-astrobiology in due course.

# Intellectual benefits of the study of astrobiology

As has often been noted previously (e.g. Connell, 2000; Race et al., 2012), the principal scientific and intellectual benefits of astrobiology arise from its inherently interdisciplinary nature. The study of astrobiology requires a grasp of, at least, astronomy, biology, biochemistry, geology, and planetary science. All undergraduate courses in astrobiology (including the one the author has taught at Birkbeck College London for many years; Birkbeck College, 2017) need to cover elements of all these different disciplines, and postgraduate and postdoctoral astrobiology researchers likewise need to be familiar with most or all of them.

From the latter half of the nineteenth century onwards these subjects have been taught in schools and universities as separate academic disciplines, with the result that the practice of science has ceased to reflect the underlying continuum of nature. Despite the undoubted advances in scientific understanding that have resulted from academic specialisation, as noted by Offerdahl (2013) there has also been an intellectual downside:

> "the structure of undergraduate curricula and courses tends to compartmentalize science into discrete disciplines that focus on particular questions rather than an integrated, interdisciplinary way of understanding the world, let alone any discussion of the societal implications of the science (Offerdahl, 2013; p.226)."

By forcing multiple scientific disciplines to interact, the study of astrobiology can stimulate, and is stimulating, a partial re-unification of the sciences and helping to move the practice of twenty-first century science back towards the more interdisciplinary outlook that prevailed in the seventeenth and eighteenth centuries (for book-length treatments of the latter see, e.g., Uglow (2002) and Holmes (2008)). By producing broad-minded scientists, familiar with multiple aspects of the natural world, the study of astrobiology therefore enriches the whole scientific enterprise. It is from this cross-fertilization of ideas that future discoveries that would not otherwise be made may be expected, and such discoveries will comprise a permanent legacy of astrobiology even if they do not include the discovery of alien life.

In addition to the (partial) re-unification of the sciences, the study of astrobiology is also stimulating much intellectual and philosophical activity outside the scope of the natural sciences, for example the numerous studies of the ethical, cultural and social implications of the discovery, or non-discovery, of life elsewhere (see, e.g., Race et al, 2012, and multiple papers in the volumes edited by Bertka (2009), Dick & Lupisella (2009), and Impey et al. (2013)). Indeed, because of this, astrobiology is well-placed to (again, at least partially) help heal the rift between science and the humanities identified by C.P. Snow in his famous 1959 Rede Lecture on the "Two Cultures" (Snow, 1959), and more recently by Wilson (1998). As noted by Kemp (2009), in an article marking the 50[th] anniversary of Snow's lecture:

> "it is the perceived need for intense specialization of any kind – in history or physics, in languages or biology – that needs to be tackled. [….] What is needed is an education that inculcates a broad mutual understanding of the nature of the various fields of research" (Kemp, 2009).

If nothing else, astrobiology can help provide just such an interdisciplinary education. Indeed, when all the aspects are considered, it seems clear that the intellectual life of humanity will be enriched on many levels merely by *searching* for life elsewhere in the universe, regardless of whether the search is successful or not.

It is also important to recognize that astrobiology comprises a research agenda that is temporally and spatially open-ended. Searching for life in the universe will take us from extreme environments on Earth, to the plains and sub-surface of Mars, the icy satellites of the giant planets, and on to the all-but-infinite variety of planets orbiting other stars. And this search will continue whether life is actually discovered in any of these environments or not. The range of entirely novel environments opened to investigation will be essentially limitless, and thus has the potential to be a never-ending source of intellectual stimulation (for a fuller discussion of this point see Crawford, 2014a).

## Cosmic and evolutionary perspectives: wider societal benefits of astrobiology

Transcending the more narrowly intellectual benefits of astrobiology, however, are a range of wider societal benefits. These arise because, as noted by Stoeger (2013), "astrobiology encourages us to expand and deepen our views of society and self." To my mind, the principal societal benefits arising from the study of astrobiology, and from its popularization to a wider public, are a consequence of the cosmic and evolutionary perspectives on human affairs that it naturally engenders.

With respect to the cosmic, perspective, it is simply not possible to consider searching for life on Mars, or on a planet around another star, without moving away from the narrow Earth-centric perspectives that dominate the social and political lives of most people most of the time. At a time when the Earth is faced with global challenges that can only be met by increased international cooperation (and arguably by developing institutions of global *governance*; see, e.g., Cabrera et al., 2011; Crawford 2014b), yet tribal nationalistic and religious ideologies are acting to fragment humanity, the promulgation of a unifying cosmic perspective on human affairs is potentially of enormous importance.

In the early years of the Space Age, the then US Ambassador to the United Nations, Adlai Stevenson, said of the world that "we can never again be a squabbling band of nations before the awful majesty of outer space" (quoted by Poole, 2008; p. 42). Yet this perspective can only influence the thinking of people if it is brought to their attention, and astrobiology can play an important role in doing so. Note that, although the cosmic perspective might be expected to intrude most forcibly into the public consciousness in the event of an actual discovery of extraterrestrial life, the search itself can also contribute. Indeed, it is only by sending spacecraft out to explore the Solar System, in part for astrobiological purposes, that we can obtain images of our own planet that show it in its true cosmic setting (Fig. 1).

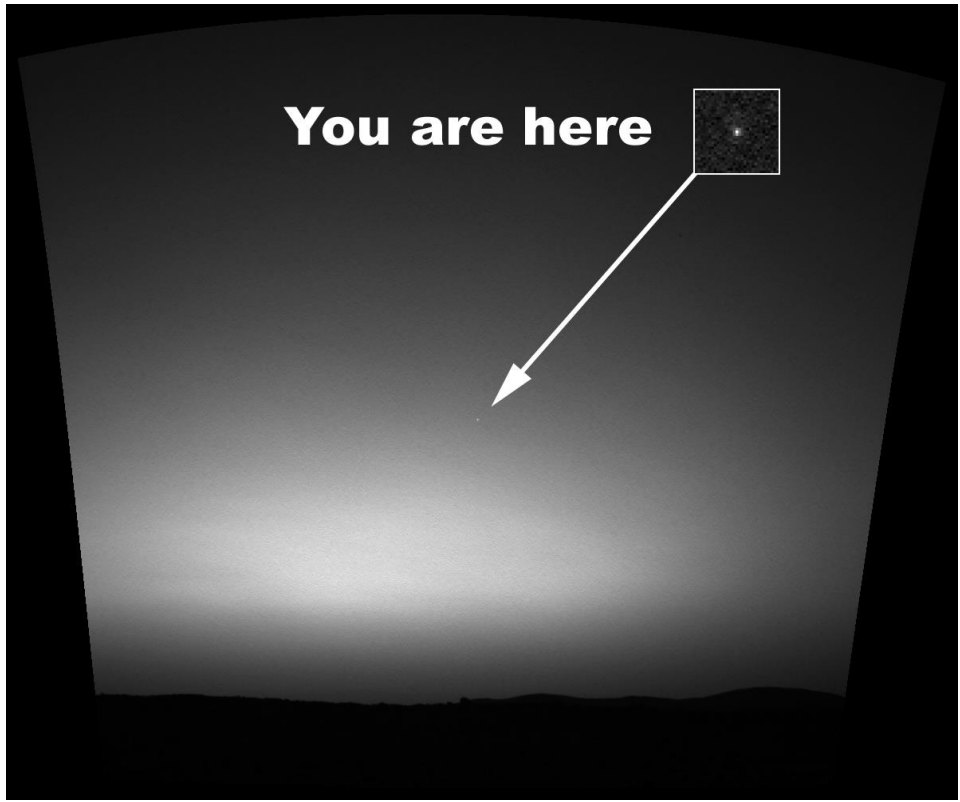

**Fig. 1.** The Earth photographed from the surface of Mars by the Mars Exploration Rover *Spirit* in March 2004. Such images powerfully reinforce a 'cosmic perspective' that can have a unifying influence on human affairs (image courtesy of NASA).

The societal benefits of this perspective have been stressed by, among others, Sagan (1994), Poole (2008), Crawford (2014b), and White (2014). Indeed, White (2014) makes the important point that:

> "it is time for the influence of space exploration on human consciousness to be seen as a legitimate justification for investing in it" (White, 2014; p. 102).

Fortunately, opportunities for developing this cosmic perspective on human affairs are likely to increase in the future, especially as a result of proposed human space missions beyond low Earth orbit, which, at least in part, will be motivated by astrobiological considerations (Crawford, 2010; ISECG, 2013).

In addition to a spatial, 'cosmic', perspective, astrobiology also provides an important temporal and evolutionary perspective on human affairs. Indeed, many university astrobiology courses, including the author's own (Birkbeck College, 2017), begin with an overview of the history of the universe, beginning with the Big Bang and moving successively through the origin of the chemical elements, the evolution of stars, galaxies, and planetary systems, the origin of life, and evolutionary history from the first cells to complex animals such as ourselves. There is a strong synergy here with the emerging discipline of 'Big History' (Christian 2004, 2017; Spier, 2010), which, following the lead of Chambers (1844, 1845), Humboldt (1845) and Wells (1925), aims to integrate human history with an evolutionary history of the cosmos.

That such a perspective has the potential to act as a unifying influence on humanity, especially by making clear our common evolutionary heritage, has long been recognized. For example, Chambers (1845) noted that such universal histories are in harmony with the aim of

> "establishing the universal brotherhood and social communion of man. And not only this, but it extends the principle of humanity to the other meaner creatures also. Life is everywhere ONE" (Chambers, 1845; p. 184, capitals in the original).

A similar observation was made by an early reviewer of the first volume of Humboldt's *Cosmos,* who noted that "[t]he individual is made to feel that he is connected, by the very nature and substance of his body, with every part of the universe" and drew the societal implication that:

> "If the world is ever to be harmonized it must be through a community of knowledge, for there is no other universal or non-exclusive principle in the nature of man" (Whelpley, 1846; p. 603).

What Whelpley apparently had in mind was a sense that humanity might be able to unify itself socially and politically if it could but agree on a common integrated worldview of the kind that Humboldt was trying to convey. More recently, the physicist David Bohm, also reflecting on the societal benefits of more holistic thinking, lamented the view of society "as a set of separately existent nations, races, or political, economic, and religious groups" because, as he noted:

> "all these parts are actually intimately related and interdependent, as aspects of an unbroken totality, which ultimately merges with the whole of existence. The idea that they are essentially separate and independent has brought about a continual series of crises throughout the whole of recorded history" (Bohm, 1996; pp. 76-77).

It hardly needs to be pointed out that the holistic worldview sought by these authors is essentially identical to that implied by the study of astrobiology and related disciplines.

There is a well-known aphorism widely attributed to Humboldt (e.g. Wilson, 2016; p. 79), and which is perhaps especially relevant at a time when narrow nationalistic and religious ideologies appear to be on the rise, to the effect that:

> "the most dangerous worldview is the worldview of those who have not viewed the world."

Humboldt was presumably thinking about the mind-broadening potential of international travel, based on his own experiences as an explorer. However, familiarity with the cosmic and evolutionary perspectives provided by astrobiology, powerfully reinforced by actual views of the Earth from space, can surely also act to broaden minds in such a way as to make the world less fragmented and dangerous. Astrobiology as a discipline can play a major part in achieving this, not least because, as stressed by Connell et al. (2000) and Offerdahl (2013), much astrobiology research is of wide public interest and often in the public eye.

## Conclusions

The cosmic and evolutionary perspectives engendered by astrobiology (and its sister discipline of Big History) deserve to be more widely appreciated in society at large. Indeed, I argue that they ought to form part of the worldview of every educated person. Crucially, many of the societal benefits of such an expanded worldview will accrue to society even if astrobiology is not successful in its ultimate aim of discovering life elsewhere in the universe. Specifically, I have argued that these benefits include:

- Stimulating the (partial) reintegration of the sciences by forcing the practitioners of different scientific disciplines to work together on highly interdisciplinary problems.

- Breaking down (albeit again only partially) some of the barriers that exist between the sciences and the humanities as the philosophical and ethical disciplines are brought to bear on issues related to the discovery (or non-discovery) of life in the universe.

- Enhancing public awareness of the 'cosmic perspective', which reveals Earth to be a very small planet adrift in the wider universe and that, to-date at least, it is the only known inhabited location in that universe.

- Enhancing public awareness of the 'evolutionary perspective', which reveals that all life on Earth is related owing to its common origin and evolutionary history.

Taken together, the last two bullet points comprise what I see as the key socio-political insight provided by the study of astrobiology and related disciplines, namely that all human beings and all human societies live on the same small planet and are related by a common evolutionary history. Although this perspective will already be obvious to readers of this journal, the point is that it is still far from being part of the world view of many of our fellow citizens and astrobiologists are uniquely placed to promulgate it to a wider public.

Moreover, to my mind, this key insight implies at least two important socio-political corollaries:

- That maintaining the continued habitability of the Earth is essential, not just for the sake of our own species, or indeed other extant species, but because

> (unless or until astrobiology itself teaches us otherwise) it may be that the entire future of life in the universe will depend upon it.

And,

- That, as a single intelligent technological species that has become dominant on a small planet of possibly unique importance to the future of life in the universe, humanity has a responsibility to develop international social and political institutions appropriate to managing the situation in which we find ourselves.

In concluding his *Outline of History*, Wells (1925; vol. II, p. 725) famously observed that: "human history becomes more and more a race between education and catastrophe." Such an observation appears especially germane to the geopolitical situation of the second decade of the twenty-first century, where apparently irrational decisions, often made by governments (and indeed by entire populations) seemingly ignorant of cosmic and evolutionary perspectives, may indeed lead our planet to catastrophe. In such an environment, the perspectives provided by the study of astrobiology may prove to be of transcendental importance, regardless of whether extraterrestrial life is ever discovered or not.

## Acknowledgements

This paper is an expanded and updated version of a talk given at a session on 'The Search for Life beyond Earth and its Implications to Society' at the 2013 European Planetary Science Congress (EPSC) at University College London (Crawford, 2013); I would like to thank Lewis Dartnell for organising that important and interesting meeting.